# Laboratory-based sticking coefficients for ices on a variety of small grains analogs

C. Laffon[1], D. Ferry[1], O. Grauby[1], Ph. Parent*[1]

[1]*Aix Marseille Univ, CNRS, CINaM, Marseille, France*

*Philippe.parent@univ-amu.fr*

**Abundances and partitioning of ices and gases produced by gas-grain chemistry are governed by adsorption and desorption on grains. Understanding astrophysical observations rely on laboratory measurements of adsorption and desorption rates on dust grains analogs. On flat surfaces, gas adsorption probabilities (or sticking coefficients) have been found close to unity for most gases[1–3]. Here we report a strong decrease of the sticking coefficients of $H_2O$ and $CO_2$ on substrates more akin to cosmic dust, such as submicrometer-sized particles of carbon and olivine, bare or covered with ice. This effect results from the local curvature of the grains, and then extends to larger grains made of aggregated small particles, such as fluffy or porous dust in more evolved media (e.g. circumstellar disks). The main astrophysical implication is that accretion rates of gases are reduced accordingly, slowing the growth of cosmic ices. Furthermore, volatile species that are not adsorbed on a grain at their freeze-out temperature will pertain in the gas phase, which will impact gas-ice partitions. We also found that thermal desorption of $H_2O$ is not modified by grains size, and thus the snowlines' temperature should be independent on the dust's size distribution.**

Grain-size distributions, from few nanometers to a micrometer, and compositions can be derived from grains' radiative properties. There may be small grains in diffuse clouds[4]; more evolved, aggregated grains in dense clouds[5]; and bigger, porous and complex grains of µm to cm-size in protoplanetary disks[6]. They include various components: polycyclic aromatic hydrocarbons (PAH), particles of carbon (graphite, amorphous carbon, organic refractory material), silicate (olivine), or/and a mixture of both. Laboratory experiments have explored many aspects of gas-grain interactions, from the initial stage of surface atom additions[7] to the more advanced stages of ice photochemistry[8]. Yet, seldom studies have been carried out on substrates close to ISM dust, such as carbon nanoparticles[9–11], nano-structured silicates[12–14], bare or mixed with ice[15], but effects on the sticking coefficient of gases were not investigated. To this end, we have performed a set of X-ray photoelectron spectroscopy (XPS) experiments to study the adsorption of $H_2O$ and $CO_2$ - which are abundant constituents of astrophysical ices - on submicrometer-sized particles of carbon and olivine, the main minerals that make up cosmic dust. We also used other materials such as $TiO_2$ or $Al_2O_3$ to help unraveling the role of chemical composition from size and morphology. Evolving with the gas exposure, the XPS signals of both the substrate and the adsorbate provide the coverage versus pressure, *i.e.* the value of the sticking coefficient S. We present in this letter the sticking coefficients of $H_2O$ and $CO_2$ on these grain analogs, bare or pre-covered with ice to model the icy mantle of cosmic dust.

As references for adsorption on plane surfaces, we used a gold foil and crystals of graphite HOPG, olivine, $TiO_2$, and $Al_2O_3$. As grain dust analogs, we used both a graphite powder with particles size of ≈2.5 µm (graphite µP) and powders of submicrometer-sized particles (µP): olivine µP (≈0.30 µm), $TiO_2$µP (≈0.24 µm), $Al_2O_3$µP (≈0.24 µm), and an organic carbon soot C1 (≈0.115 µm) produced by a soot laboratory generator. To mimic the smallest dust grains, we used nanoparticles consisted either of a graphitic carbon soot C2 (≈0.025 μm) produced by the same generator, or a carbon soot produced by a candle CS (≈0.025 μm), and the same soot after an oxygen plasma oxidation, called Ox-CS. The grain size corresponds to the mode of the particle size distribution, *i.e.* the most commonly found diameter. Fig. 1 presents a selection of electron microscopy images of the substrates; more details on the samples are given in Methods.

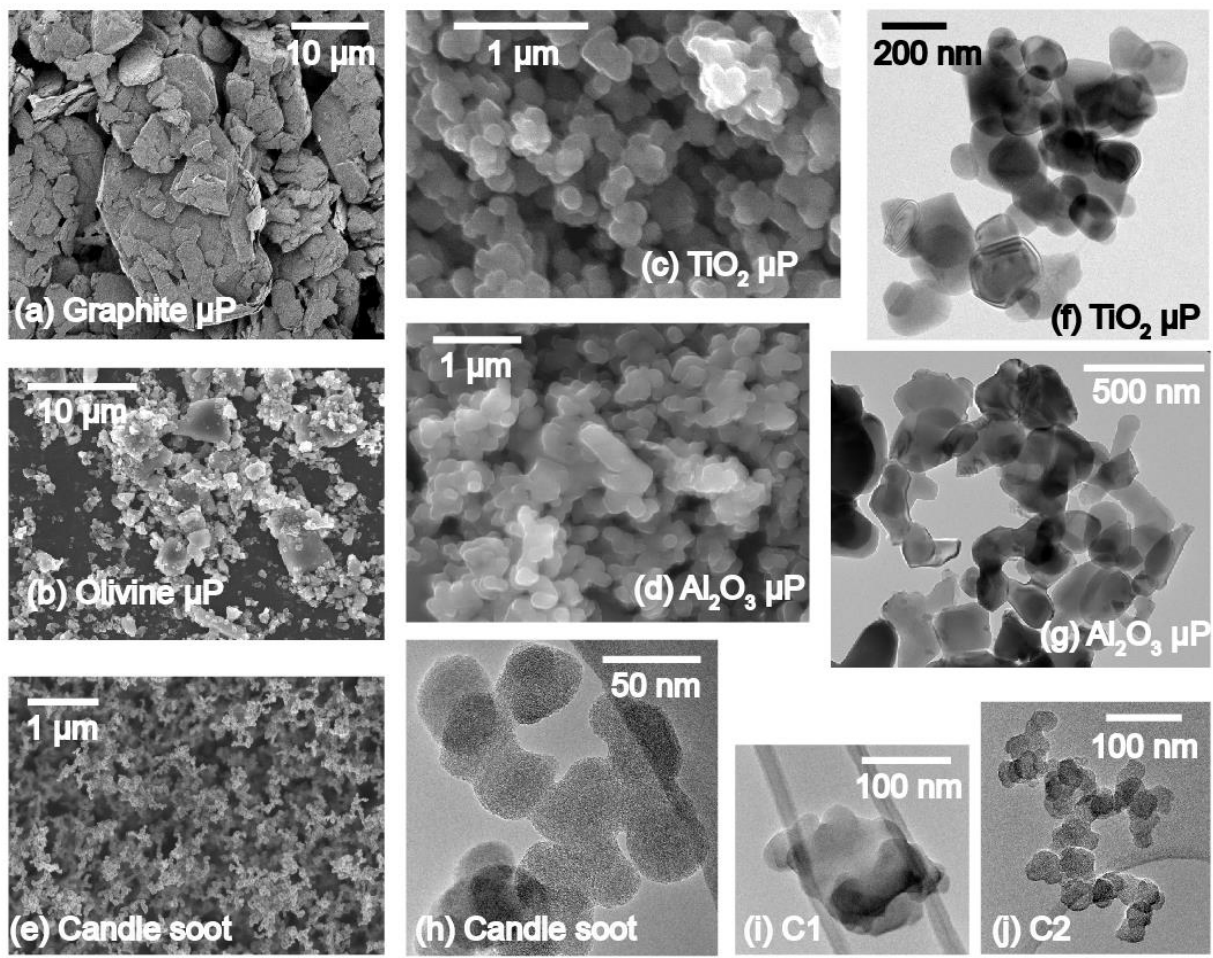

**Fig. 1 Selected electron microscopy images of the substrates**. Scanning electron microscopy images of the dust grain analogs : (a) graphite μP; (b) olivine μP; (c) TiO$_2$μP; (d) Al$_2$O$_3$μP;(e) candle soot. Transmission electron microscopy images of (f) TiO$_2$μP; (g) Al$_2$O$_3$μP; (h) Candle soot; (i) C1 soot; (j) C2 soot**.**

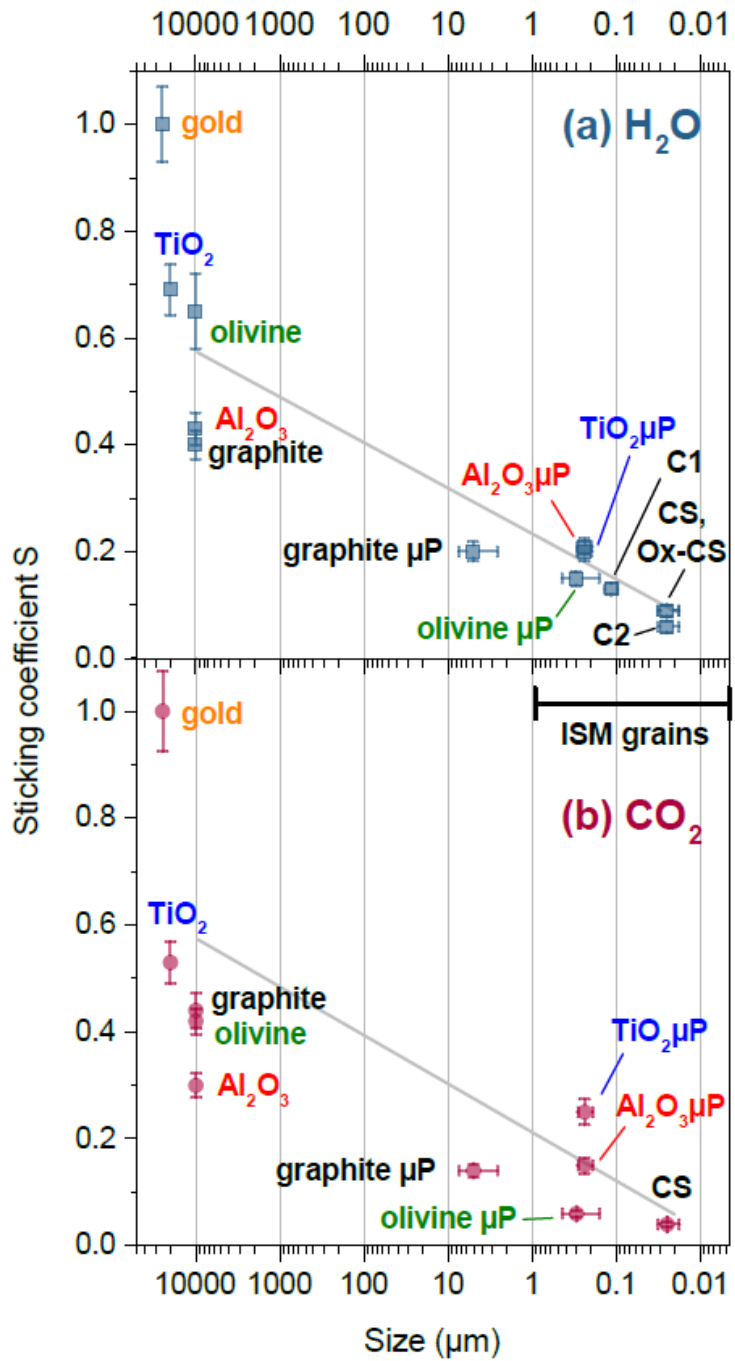

**Fig.2 Sticking coefficients of H₂O and CO₂ at 20 K, as a function of the substrate size (log scale), and for different substrates**. Mean values of sticking coefficients S of (a) H₂O and (b) CO₂ at 20 K plotted as a function of the substrate size, and for different materials to which colors are assigned (orange: gold; black: carbon; blue: TiO₂; green: olivine; red: Al₂O₃); lines are eyes guides. The horizontal bar indicates the diameter range of ISM grains. Errors on S are estimated between ±5% and ±10% of standard deviation, depending on the substrate and the adsorbate (see Methods and Supplementary Information 3). Error bars on the size are the width of the particle size distribution.

The substrates are cooled but the gases are not, which raises the question of whether our results can be extrapolated to the ISM, wich is discussed in Methods. Fig. 2 presents the mean sticking coefficients S of $H_2O$ at 20 K on all substrates (top), and for $CO_2$ at 20 K on fewer substrates (bottom). Further experiments on $H_2O$ at 80 K led to values close to those measured at 20 K (see Supplementary Information 5). The sticking coefficients are graphically presented as a function of the substrate size (note the logarithmic scale). S values for increasing exposures are also listed Table 1 (the mean values are plotted in Fig. 2). In Table 1, changes of S with exposure are related to the evolution of the growth mode with the coverage (see Supplementary Information 2). In Fig. 2, the horizontal bar indicates the diameter range of ISM grains in dust models consistent with astronomical observations[16]. Decreasing the grain size causes a decrease in the sticking coefficient of $H_2O$ and $CO_2$. The highest values of S are obtained for adsorption on plane surfaces. They are quite dispersed ($0.40 < S < 1$ for $H_2O$, and $0.30 < S < 1$ for $CO_2$) due to different chemical structures leading to different densities of adsorption sites. It is worth mentioning that S differs from 1 on surfaces commonly used as grain analogs in laboratory astrochemistry: $S = 0.65$ on olivine and 0.40 on graphite HOPG, in agreement with Refs.[17,18]. The values of S are substantially lower on submicrometer-sized grains (C1, olivine µP, $Al_2O_3$µP, $TiO_2$µP) and micrometer-sized grains (graphite µP): $0.13 < S < 0.21$ for $H_2O$ and $0.06 < S < 0.25$ for $CO_2$. There is also a certain dispersion of the values of S, showing that adsorption is still influenced by chemical composition, as on plane surfaces. Finally, the values of S are the lowest for adsorption onto CS, C2 and Ox-CS carbon soot nanoparticles: $0.06 < S < 0.09$ for $H_2O$, and 0.04 for $CO_2$. There is no difference after a strong oxidation of CS ($S = 0.09$ for CS and Ox-CS) while oxidized carbon functions should favor water adsorption by hydrogen-bonding. This indicates that chemical composition of such nano-sized particles has little or no influence on adsorption. Furthermore, it is important to note that thermal desorption experiments of $H_2O$ indicates that desorption temperature is not dependent on the substrate's size (see Supplementary Information 4).

**Table 1**. Sticking coefficients of $H_2O$ and $CO_2$ at increasing exposures, and the corresponding mean value S plotted Fig. 2 and Fig. 3 (error estimations are explained SI-3).

| $H_2O$ at 20K | Size | 0.1 L | 0.25L | 0.5 L | 0.75L | 1 L | S |
|---|---|---|---|---|---|---|---|
| Gold | 1.0 cm | 1 | 1 | 1 | 1 | 1 | 1 ± 0.07 |
| $TiO_2$ | 1.0 cm | 0.56 | 0.63 | 0.71 | 0.78 | 0.77 | 0.69 ± 0.05 |
| Olivine | 1.0 cm | 0.56 | 0.52 | 0.67 | 0.81 | 0.71 | 0.65 ± 0.07 |
| $Al_2O_3$ | 1.0 cm | 0.39 | 0.37 | 0.40 | 0.47 | 0.53 | 0.43 ± 0.07 |
| Graphite HOPG | 1.0 cm | 0.30 | 0.37 | 0.42 | 0.45 | 0.46 | 0.40 ± 0.03 |
| Graphite µP | 5.0 µm | 0.22 | 0.19 | 0.20 | 0.20 | 0.20 | 0.20 ± 0.02 |
| Olivine µP | 0.30 µm | 0.18 | 0.15 | 0.14 | 0.13 | 0.13 | 0.15 ± 0.01 |
| $Al_2O_3$ µP | 0.24 µm | 0.18 | 0.17 | 0.20 | 0.17 | 0.24 | 0.20 ± 0.02 |
| $TiO_2$ µP | 0.24 µm | 0.15 | 0.17 | 0.18 | 0.18 | 0.17 | 0.21 ± 0.01 |
| C1 | 0.115 µm | 0.11 | 0.12 | 0.14 | 0.14 | 0.14 | 0.13 ± 0.01 |
| Candle Soot | 0.025 µm | 0.09 | 0.08 | 0.08 | 0.09 | 0.10 | 0.09 ± 0.01 |
| OX-CS | 0.025 µm | 0.08 | 0.08 | 0.09 | 0.09 | 0.10 | 0.09 ± 0.01 |
| C2 | 0.025 µm | 0.06 | 0.06 | 0.06 | 0.05 | 0.06 | 0.06 ± 0.01 |
| | | | | | | | |
| $CO_2$ at 20 K | Size | 0.1 L | 0.25 L | 0.5 L | 0.75 L | 1 L | S |
| Gold | 1.0 cm | 1 | 1 | 1 | 1 | 1 | 1 ± 0.07 |
| $TiO_2$ | 1.0 cm | 0.67 | 0.48 | 0.50 | 0.48 | 0.50 | 0.53 ± 0.04 |
| Olivine | 1.0 cm | 0.42 | 0.55 | 0.37 | 0.36 | 0.42 | 0.42 ± 0.02 |
| $Al_2O_3$ | 1.0 cm | 0.28 | 0.26 | 0.29 | 0.33 | 0.36 | 0.30 ± 0.02 |
| Graphite HOPG | 1.0 cm | 0.45 | 0.45 | 0.42 | 0.40 | 0.46 | 0.44 ± 0.03 |
| Graphite µP | 5.0 µm | 0.19 | 0.15 | 0.09 | 0.09 | 0.10 | 0.12 ± 0.01 |
| Olivine µP | 0.30 µm | 0.07 | 0.07 | 0.07 | 0.05 | 0.05 | 0.06 ± 0.01 |
| $Al_2O_3$ µP | 0.24 µm | 0.19 | 0.15 | 0.14 | 0.14 | 0.14 | 0.15 ± 0.01 |
| TiO$_2$ µP | 0.24 µm | 0.30 | 0.27 | 0.24 | 0.21 | 0.23 | 0.25 ± 0.02 |
| Candle Soot | 0.025 µm | 0.04 | 0.05 | 0.04 | 0.04 | 0.05 | 0.040 ± 0.003 |
| | | | | | | | |
| **CO$_2$ at 20 K/ASW** | **Size** | **0.1 L** | **0.25 L** | **0.5 L** | **0.75 L** | **1 L** | **S** |
| Gold | 1.0 cm | 1 | 1 | 1 | 1 | 1 | 1± 0.07 |
| TiO$_2$ | 1.0 cm | 0.89 | 0.90 | 0.93 | 0.96 | 0.97 | 0.93 ± 0.07 |
| Al$_2$O$_3$ | 1.0 cm | 0.43 | 0.47 | 0.61 | 0.74 | 0.87 | 0.62 ± 0.05 |
| Graphite HOPG | 1.0 cm | 1.00 | 0.81 | 0.77 | 0.74 | 0.75 | 0.81 ± 0.06 |
| Al$_2$O$_3$ µP | 0.24 µm | 0.49 | 0.43 | 0.46 | 0.42 | 0.43 | 0.45 ± 0.03 |
| TiO$_2$ µP | 0.24 µm | 0.47 | 0.49 | 0.51 | 0.49 | 0.48 | 0.49 ± 0.05 |
| Candle Soot | 0.025 µm | 0.03 | 0.04 | 0.05 | 0.05 | 0.04 | 0.040 ± 0.003 |
| | | | | | | | |
| **CO$_2$ at 20 K/p-ASW** | **Size** | **0.1 L** | **0.25 L** | **0.5 L** | **0.75 L** | **1 L** | **S** |
| Gold | 1.0 cm | 1 | 1 | 1 | 1 | 1 | 1 ± 0.07 |
| TiO$_2$ | 1.0 cm | 0.71 | 0.65 | 0.78 | 0.75 | 0.83 | 0.74 ± 0.05 |
| Al$_2$O$_3$ µP | 0.24 µm | 0.23 | 0.27 | 0.29 | 0.32 | 0.30 | 0.28 ± 0.03 |
| TiO$_2$ µP | 0.24 µm | 0.32 | 0.30 | 0.34 | 0.35 | 0.37 | 0.34 ± 0.03 |
| Candle soot | 0.025 µm | 0.03 | 0.04 | 0.04 | 0.04 | 0.04 | 0.040 ± 0.003 |

To model condensation on the icy mantle of cosmic dust, we have adsorbed $CO_2$ on grains pre-covered with a thin layer of ice preserving the morphology of the underlying grain (see Methods). As the degree of porosity of cosmic ices is unknown, two kinds of ice were deposited: a porous amorphous ice (p-ASW, for porous Amorphous Solid Water) and a non-porous amorphous ice (ASW). It can be thus determined whether adsorption depends on the ice porosity. Fig. 3 presents the sticking coefficients of $CO_2$ at 20 K on a selection of bare and icy substrates; for the sake of clarity, only the TiO$_2$ plane surface is presented (further substrates are listed in Table 1). Ice substantially increases the sticking of $CO_2$, from S=0.53 on bare TiO$_2$ to S=0.74 on p-ASW/TiO$_2$, and S=0.93 on ASW/TiO$_2$, a value in good agreement with He et al.[3]. This is also consistent with the work of Noble et al.[19] who reported that $CO_2$ better sticks on non-porous ice. Adsorption of $CO_2$ is also enhanced by ice on submicrometer particles, from S=0.25 and 0.15 ($TiO_2\mu P$ and $Al_2O_3\mu P$) to 0.34 and 0.28 (p-ASW/$TiO_2\mu P$, p-ASW/ $Al_2O_3\mu P$), and to 0.49 and 0.45 (ASW/$TiO_2\mu P$, ASW/$Al_2O_3\mu P$). However, the values of S are not as high as those measured on ice surfaces deposited on flat $TiO_2$, indicating that the size of the icy grain still influences adsorption. Last, sticking coefficients of $CO_2$ on bare and ice-covered carbon nanoparticles (CS) are extremely low and identical (S=0.04 in both cases). This adds evidence that surface chemistry plays very little role in the case of nanoparticles.

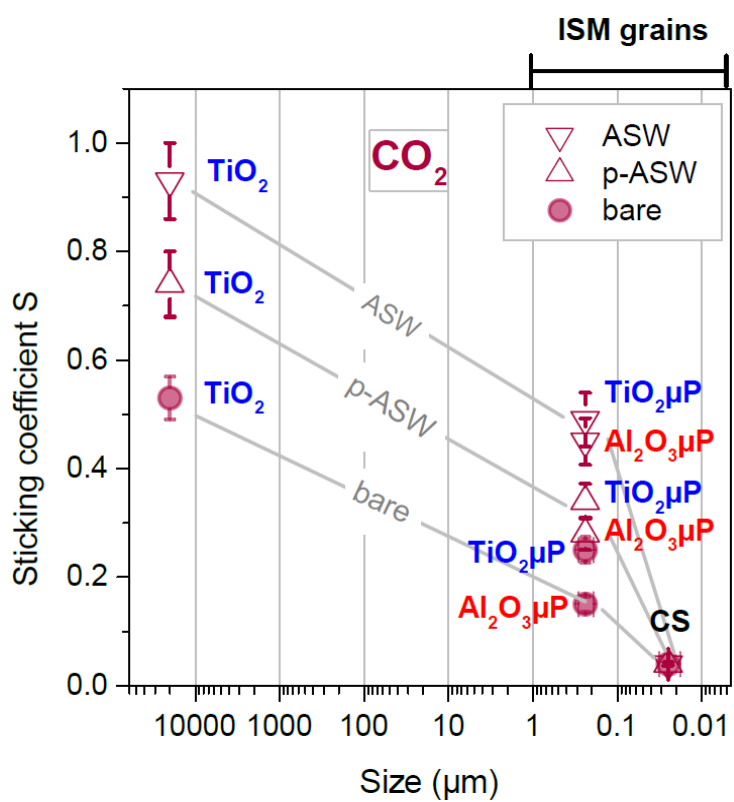

**Fig.3 Sticking coefficients of $CO_2$ on $H_2O$ ice at 20 K, as a function of the substrate size (log scale), and for different substrates**. Mean values of sticking coefficients S of $CO_2$ at 20 K on a selection of substrates either bare or pre-covered with amorphous ice (ASW) or porous amorphous ice (p-ASW), plotted as a function of the substrate size, and for different materials to which colors are assigned (black: carbon; blue: $TiO_2$; red: $Al_2O_3$); lines are eyes guides. Errors on S are estimated between ±5% and ±10% of standard deviation, depending on the substrate and the adsorbate (see Methods and Supplementary Information 3). Error bars on the size are the width of the particle size distribution.

In the Earth's atmosphere, it is well established that the ability of solid aerosols to condense water depends on their chemistry[20], but the size of the particle also matters. Indeed, increasing curvature weakens the attractive forces between the adsorbed water molecules, and the equilibrium vapor pressure of a gas is always higher at convex surfaces than at flat surfaces; this reduces or hinders the condensation of water onto small aerosol particles. This effect, classically described by the Kelvin's equation, becomes crucial at nanometer scales[21]. It causes the poor cloud-nucleating ability of small atmospheric aerosols, and is independent of the composition for micrometer-sized particles[22]. Regarding ice, molecular simulations indicate that heterogeneous nucleation is indeed less efficient at surfaces whose convex curvature is not negligible at the molecular scale[23]. For instance, CO interacts more strongly in surface valleys than on hills of the amorphous ice surface[24]. Simulations also show that on rough ice nanoparticles, molecules physisorb in crevices and indentations rather than on bumps and protrusions[25]. For an amorphous ice cluster, quantum-mechanical simulations show that the low-lying potential minima are indeed located in cavities[26]. This is the microscopic rendering of Kelvin's equation, which also states that concave surfaces facilitate nucleation by increasing the number of van der Waals interactions with the wall of the confined space, a phenomenon called capillary condensation. On nanoparticles such as soot, morphology is dominated by highly convex surfaces where gases will then condense less efficiently. On such small particles, a modification of the surface chemistry by oxidation or ice coating does not make any difference, because the grain morphology remains unchanged. For larger, submicrometer-sized particles ($TiO_2\mu P$, $Al_2O_3\mu P$, olivine $\mu P$), strongly convex areas are necessarily fewer; their morphology is mostly made up of slightly curved surfaces, and adsorption is mainly controlled by chemistry. Last, the sticking coefficient being governed by the local features of the grain, it is not determined at the grain's overall scale, but at the particles that make it up. Our results therefore apply to various cosmic environments, whether dust consists of single grains (diffuse ISM) or small grains coagulated into larger aggregates (dense clouds, protoplanetary disks).

One main general astrophysical implication is that accretion rates of gases on grains will be reduced, and cosmic ices will grow at slower rates[27]. Furthermore, since not all species adsorb when hitting a grain at their freezing temperature, they can remain in the gas phase instead of forming the expected ices. This must be considered to model the partition between gases and ices in cold regions, along with possible non-thermal desorption processes that reallocate molecules in the gas phase, when present. Until experimental values are determined, some recommendations for the values of S can be provided to reassess astrochemical models of interstellar and circumstellar environments. For highly curved, small grains (≤ 0.025 μm in diameter), or bigger aggregate made of such grains, chemical composition plays no role, and the sticking coefficients are below 0.1. For less curved, larger grains (≈ 0.2 μm), or bigger aggregate of such grains, adsorption can vary with chemical composition and curvature; the sticking coefficients range from 0.1 to 0.3 on bare grains, and from 0.3 to 0.5 on icy grains (this size range encompasses the 0.2 μm canonical grain commonly used in astrochemical models). If we extend the results obtained at 20 K on $H_2O$ and $CO_2$ (whose direct condensation on grains is unlikely to occur in the ISM) to species of astrophysical interest like CO, C, O, N, S for the ISM[7], and $N_2$, $NH_3$, $CH_3OH$, $CH_4$ for protoplanetary regions[28], we would recommend using the sticking coefficients measured for $H_2O$ regarding those having the ability to form a hydrogen-bond, like $NH_3$, or being highly polar like $CH_3OH$. For non-polar species like CO, $CH_4$, $N_2$, C, O, N, S, the sticking coefficients of $CO_2$ might be a better choice. Regarding H, D, $H_2$ and D$_2$, their sticking coefficients might be substantially smaller on small grains than on reference surfaces commonly used. However, not to mention the effects of gas temperature (see Methods), their masses are too different from those of CO$_2$ and H$_2$O to use the values of S measured in this work. As an example of astrophysical implication, taking S(CO)≈0.5 on icy grain will allow a better quantification of CO in outer envelopes of pre-stellar objects. For the pre-stellar core L1689B, this halves the CO freeze-out rate and leads to time-scales closer to the nominal free-fall time, in better agreement with the observed abundance[29]. Finally, water playing a central role in planet-forming region (disk composition, coagulation)[30], assessing realistic conditions for direct condensation and sublimation of H$_2$O on small grains is important for interpreting water abundances. Our study shows that while grain size impacts the condensation process, it has no effect on the desorption temperature (see Supplementary Information 4). An important consequence is that the isotherm of the water snowline in planetary disks will not be dependent of the size distribution of the grains.

**Methods**

**Samples.** As plane, cm-sized substrates, we used (1) a gold foil of 10 x 10 x 0.5 mm$^3$ (99.95%, Alfa Aesar); (2) a highly oriented pyrolytic graphite of 10 x 10 x 1 mm$^3$ (HOPG, Alfa Aesar); (3) a crystal of TiO$_2$ (rutile) of 10 x 10 x 2 mm$^3$, (4) a crystal of Al$_2$O$_3$ (corindon) of 10 x 10 x 5 mm$^3$, and (5) a crystal of peridot of 10 x 10 x 1 mm$^3$. Peridot is a gem variety of a natural Mg-rich San Carlos olivine, of composition (Mg$_{0.62}$,Fe$_{0.025}$)$_2$SiO$_4$ as determined by X-ray photoelectron spectroscopy (XPS). Samples (3)-(5) were provided by the collection of mineralogy of Aix-Marseille University, and cut and polished mechanically. Powders with particles of various sizes were used as grain analogs: (6) graphite µP made graphite grains of ≈ 5.0 (±2.4) µm in diameter, sometimes aggregated in bigger grains of 20-30 µm; (7) olivine µP powder obtained after fine grinding of an olivine crystal, leading to grains of 0.30 (±0.14) µm in diameter, mixed with some bigger grains of few µm; (8) $TiO_2$µP (anatase) made of slightly faceted round grains of 0.24 (±0.05) µm in diameter, and (9) $Al_2O_3$µP (corindon) made of grains of 0.24 (±0.05) µm in diameter, obtained after drying of an ultra-pure alumina suspension (99.98 %, Presi). We also used carbon soot : (10) a candle soot (CS) emitted from a pure paraffin candle, made of fractal aggregates of graphitic carbon nanoparticles (≈90 %, the rest being aliphatic and aromatic organic hydrocarbons) with d=0.025 µm (*i.e.* 25 (±15) nm), exhibiting an imperfect graphitic structure (turbostratic disorder); (11, 12) two soot samples generated by a MiniCAST generator under two different combustion conditions, called C1 and C2[31]. C1 is an organic carbon soot (called CAST3 in Ref.[31], containing 38% of aliphatic and aromatic hydrocarbons, and 62 % of graphitic carbon), consisting in almost indistinct carbon nanoparticles forming aggregates of 0.115 (±0.02) µm in diameter. C2 is a graphitic carbon soot (94 % of graphitic carbon) made of fractal aggregates of well distinct carbon nanoparticles of 0.025 µm (*i.e.* 25 (±15) nm) with a turbostratic disordered structure (called CAST1 in Ref.[31]). XPS allowed determining surface oxidation ratios [O]/[C] of 9 %, 4 % and 3 % for C1, C2 and CS, respectively. No other atomic species, like nitrogen, were detected. The crystal structure of all samples was obtained by X-ray diffraction carried out with a PANalytical X'Pert diffractometer (not presented). The grains were also studied with electron microscopy, using a Jeol JSM-6340F microscope in scanning mode (SEM) and a Jeol JEM-2010 microscope in transmission mode (TEM). The particles size distributions were determined from electron microscopy images by using the open source image processing software ImageJ

(https://imagej.nih.gov/ij). SEM or TEM images displaying a large number of particles are loaded in the software. The spatial scale is then calibrated using the scale provided by the microscopes (horizontal bars on the images). The area selection tool of the software allows to manually define areas for several hundreds of particles. For each selected area, the software provides the area, the perimeter, and the Feret's equivalent diameter, taken as the particle's sizes. Statistics including the mean size, standard deviation, are then calculated.

The substrates were also pre-covered with two kinds of ice. In a first set of experiments, ice was condensed at 20 K to form a microporous amorphous ice film (p-ASW). In a second set of experiments, ice was condensed at 20 K, then annealed at 110 K to form a non-porous amorphous ice film (ASW)[32,33]. The thickness of the ice layers can be estimated by multiplying the water exposure by the sticking coefficient of $H_2O$ on each of the considered substrates. We obtained 14 monolayers (ML) on the plane $TiO_2$ surface, 5 ML on $TiO_2\mu P$ and 4 ML on $Al_2O_3\mu P$ particles, and 2 ML on CS. The ice layers are thin enough to avoid any significant increase of the particle's size. We found that the XPS signal from the substrates becomes negligible, indicating that they are fully covered by the ice layers.

**X-ray photoelectron spectroscopy (XPS).** Ethanol solutions of the powders were drop casted on the XPS sample holders and dried to form a thin film. The MiniCAST samples were deposited by thermophoresis onto gold-coated silicon windows (UQG optics). These windows, like the flat references (graphite, $TiO_2$, $Al_2O_3$ and olivine), were glued on the sample holders using a thermally conductive silver paint. Candle soot was directly deposited on a sample holder by passing it in the flame. Another sample of candle soot (called Ox-CS) was oxidized under ultra-high vacuum (UHV) with an $O_2$ plasma from a microwave source (Tectra GmbH). XPS showed that the oxygen concentration at the surface raised 7-fold, from 3 % to 21 %, mainly due to ketone and carbonyl functions. Graphite HOPG was cleaved before its introduction under UHV, while $TiO_2$, $Al_2O_3$ and olivine crystals were cleaned by sputtering $Ar^+$ at 3 kV for 5 min under UHV at room temperature. The $TiO_2\mu P$, $Al_2O_3\mu P$ and olivine µP samples were sputtered at the same time. All samples submitted to sputtering should be amorphous over a depth of few nm from the surface. The substrates were cooled down overnight to 20±1 K, and the proper thermalization of their surface was systematically verified by measuring with XPS the desorption temperature of an ice layer (see Supplementary Information 4). The bare substrates were exposed to $H_2O$ or $CO_2$ via background dosing (see Supplementary Information 1).

As said previously, the gases are not cooled, which raises the question of whether our results can be extrapolated to the ISM. For $H_2$, it is known that gas temperature significatively impacts the sticking coefficients[2], and to a lesser extent for $D_2$ since surface thermalization is more efficient for higher masses[34]. Gas temperatures of 10-20 K are then required to obtain realistic experimental values of the sticking coefficients of $H_2$, $D_2$, H and D. For heavier species like $H_2O$ or $CO_2$, the ISM temperatures are in practice impossible to obtain. However, their high mass should favor their thermalization[35]. This is confirmed by a rare experiment that showed that the thermal energy of CO from 30 K to 600 K is totally absorbed by a tungsten surface maintained at 20 K[36]. This also agrees with calculations predicting the full thermalization of CO at 300 K landing on water ice at 40 K[24]. For thermal CO as well as for Ar[35], the number of collisions is 2-3 before sticking, slightly increasing with the incident energy of the gas. For hydrogen at 300 K on crystalline ice at 10 K, sticking occurs a few ps after collision, i.e. ≈ 0.1 nm[37], in agreement with the calculations of Buch et al. for hydrogen at 200 K on 96 K ice cluster[38]. Thus, gas species, even the lightest, do not travel a long distance after collision on a surface held at low temperature. A further experimental evidence is the fact that background-deposition of $H_2O$ at 300K on cold substrates held below 90 K results in the growth of porous ASW film at the atomic scale, a consequence of the limited molecular mobility. Therefore, if the grain size is larger than the few interatomic distances necessary for a complete thermalization of the gas during the sticking trajectory, there is no impact of the size on the sticking probability due to this dynamic aspect, provided the surface is cold and the gas is at room temperature and not too hot. Then, we assume that the sticking coefficients of heavy species at ambient gas temperature might not be very different from those at the colder gas temperatures of the ISM.

The XPS data were recorded using a Resolve 120 hemispherical electron analyzer (PSP Vacuum), and an unmonochromatized X-ray source (Mg $K_\alpha$ at 1253.6 eV, PSP Vacuum) operated at 100 W at an angle of 115° with respect to the analyzer axis, and a detection angle of 55° between the surface normal and the analyzer axis. At this angle, called the "magic angle", the influence of roughness on the signal intensity is null or very limited (see Supplementary Information 2). During the experiments, neither the substrates nor the condensed $H_2O$ or $CO_2$ layers show any detectable radiation damage, such as chemical changes. We observed no temperature increase of the substrates due to the proximity of the X-ray tube, thanks to the low power used, the large source-sample distance (2 cm), and the adsorption by the cryostat of the thermal radiation emitted by the X-ray tube. The sticking coefficient S on each substrate is graphically determined from the XPS intensities (see Supplementary Information 2) by finding the exposure necessary to obtain the same XPS intensity than on gold, where S=1 by calibration. Uncertainty in the values of S is estimated between ±5% and ±10% of standard deviation, considering the accuracy of the exposure, error due to roughness, and the accuracy of the XPS data analysis (±5%) (see Supplementary Information 3).

**Data availability**

A text version of Table 1 is available at https://figshare.com/articles/dataset/_/13274840. All the datasets generated and analyzed during the current study are available from the corresponding author on reasonable request.

Correspondence and requests for materials should be addressed to philippe.parent@univ-amu.fr (Ph. PARENT).

**Acknowledgements**

The authors are grateful to I. Marhaba for her help during the preliminary XPS experiments, and to F.-X Ouf for providing the MiniCAST samples.

**Authors contributions**

C. L. and Ph. P. conceived, performed, analyzed and interpret the XPS experiments. O. G. provided and prepared most of the samples, and carried out the X-ray diffractions experiments. D. F. and O. G. performed and analyzed the SEM and TEM images. Ph. P. wrote the manuscript. All authors contributed ideas to this letter.

**Competing interest**

The authors declare no competing interests.

# Laboratory-based sticking coefficients for ices on a variety of small grains analogs

C. Laffon[1], D. Ferry[1], O. Grauby[1], Ph. Parent[1]

[1]*Aix Marseille Univ, CNRS, CINaM, Marseille, France*

**Supplementary Information 1. Apparatus, exposures and pressure calibration**

Supplementary Figure 1 presents a schematic description of the experiment used in this study. After preparation of the substrates described in the Method section, the sample holder is first introduced in a fast entry lock (#1) and pumped overnight to reach secondary vaccum ($10^{-7}$ Torr). It is then transferred in a preparation chamber (#2) pumped under ultra-high vacuum (UHV, $5.10^{-10}$ Torr), and picked up again for a second transfer to the analysis chamber (#3) that is evacuated with a turbomolecular pump (Pfeiffer Hipace 700), a titanium sublimation pump, and a cryo-pump (Oerlikon Coolvac 2000 CL). The base pressure is $5.10^{-10}$ Torr and even lower when the cryo-pump is in operation, *i.e.* below the low-pressure limit of the Bayard Alpert (BA) gauge fitted to the chamber (*i.e.* < $1.5\ 10^{-10}$ Torr). This chamber is equipped with the instruments for XPS, consisting in an X-ray tube and a hemispherical electron analyzer. It is also equipped with a surface infrared FTIR spectrometer in reflection-absorption (RAIRS) mode [1], not sketched. It can be used simultaneously to XPS, but in this work it was only used to calibrate the dosing pressure (see below). The sample is set on a 4-axis manipulator (θ,x,y,z) attached to a temperature-controlled closed-cycle helium refrigerator. If necessary, the sample can be lifted prior to cooling to an auxiliary UHV chamber (#4) for ion sputtering, using a cold-cathode ion gun (Thermo VG). This chamber, whose base pressure of $5.10^{-10}$ Torr is achieved using a turbomolecular pump (Pfeiffer Hipace 400), can be isolated from the XPS analysis chamber by a gate valve.

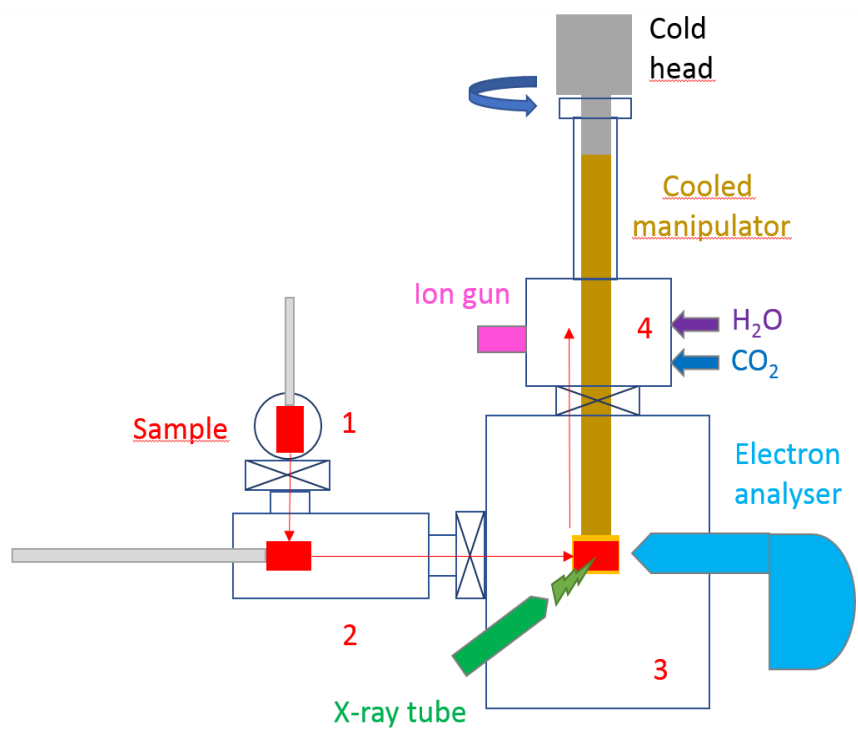

**Supplementary Figure 1** Schematic description of the experiment. The sample is first pumped in a fast entry lock (#1), then transferred via a preparation chamber (#2) onto the cooled 4-axis manipulator of the main analysis chamber (#3), where XPS (and RAIRS) analysis are carried out. Gas exposures on the cooled samples are achieved through separated leak valves in an auxiliary chamber (#4), where the sample is translated and isolated to the main chamber to avoid its contamination. When necessary, ion sputtering for surface cleaning can be achieved (at room temperature) in chamber 4 prior to the dosing experiments.

For an adsorption experiment, the manipulator is cooled and the sample is thermalized overnight at 20 ± 1K; the temperature is measured with a silicon diode (Lakeshore DT-470-DI) clamped on the cold finger. Once the XPS spectrum of the bare surface has been recorded in the analysis chamber 3, the manipulator is lifted back to the auxiliary UHV chamber 4 for dosing. Cooling the manipulator provides an extra pumping, and in this case the base pressure in chamber 4 is < 1.5 $10^{-10}$ Torr, also below the low-pressure limit of the other BA gauge fitted to this chamber.

The sample is then exposed in chamber 4 to a continuous gas flux, either $H_2O$ (from outgassed ultrapure water) or $CO_2$ vapor (99.995 %, Linde) through separated leak valves. During exposures, the gate valve between chambers 3 and 4 is closed to avoid any contamination of chamber 3. A controlled equilibrium pressure of few $10^{-8}$ Torr is maintained in chamber 4 as long as necessary for the desired exposure. Once the dosing is achieved and the base pressure in chamber 4 recovered, the manipulator is translated back to the analysis chamber 3 for the XPS recording of the covered substrate. It is then lifted up back to chamber 4 for the next dosing, etc.

The leak valves being located at the back of the sample, condensation occurs from the background pressure. The pressure being constant and isotropic, the gas molecules move randomly, and there is no net flow in a particular direction. This ensures that any open surface of any orientation located at the substrate-vacuum interface is exposed to the same mass flux during the same time, *i.e.* is subjected to the same exposure. As a surface-sensitive method, XPS will only probe these open surfaces. The dosing pressure is measured by the BA gauge of chamber 4, calibrated using the infrared spectrometer. For pressure calibration, we monitored the growth of the 3280 $cm^{-1}$ band of a pure $H_2O$ ice film, and that of the 2343 $cm^{-1}$ band of a pure $CO_2$ ice film deposited at 20 K onto a gold substrate, as function of the $H_2O$ or $CO_2$ pressures indicated by the BA gauge. The infrared band intensities allow calculating the column density of the deposited layers[2], assuming a sticking coefficient of 1 during the deposition, then providing the true molecular flux to which the substrate was submitted. This enables to determine the true dosing pressure, and therefore to correct the gauge reading by the appropriate factor for each gas.

Furthermore, we found that the condensation rates of $H_2O$ or $CO_2$ on gold (*i.e.* at submonolayer regime) is similar to that in the multilayer regime, where S=1 as stated in Ref. 2. The value of the sticking coefficient of $H_2O$ and $CO_2$ on gold at 20 K is thus set at 1.

The exposure $E$ of a surface to a gas for a dosing time $t$, usually expressed in Langmuir units ($1\,\mathrm{L} = 10^{-6}\,\mathrm{Torr.s}$), is related to the dosing pressure $P$ by the Hertz-Knudsen equation :

$$ E = \frac{Pt}{\sqrt{2\pi m k_B T}} \tag{S1} $$

Where $k_B$ is the Boltzmann constant, $m$ the molecular mass and $T$ the gas temperature. In this study, $T = 300\,\mathrm{K}$ as the substrate is cooled but not the gas. The resulting coverage is obtained by multiplying $E$ by the sticking coefficient $S$, and can be expressed in monolayer equivalent (1 ML= $10^{15}$ atoms.cm$^{-1}$).

### Supplementary Information 2. Determination of the sticking coefficient from the XPS analysis

The principle of our method is illustrated Supplementary Figure 2 with the candle soot (CS) film, exposed at T=20 K to $H_2O$ (a) and $CO_2$ (b). The evolutions of the C1s (282-294 eV) and O1s (530-540 eV) peaks are presented (unprocessed) as a function of the $H_2O$ and $CO_2$ exposures (in L). The intensities related to condensed $H_2O$ (O1s at ∼ 534 eV) or $CO_2$ (C1s at ∼ 292 eV and O1s at ∼ 536.0 eV) increase

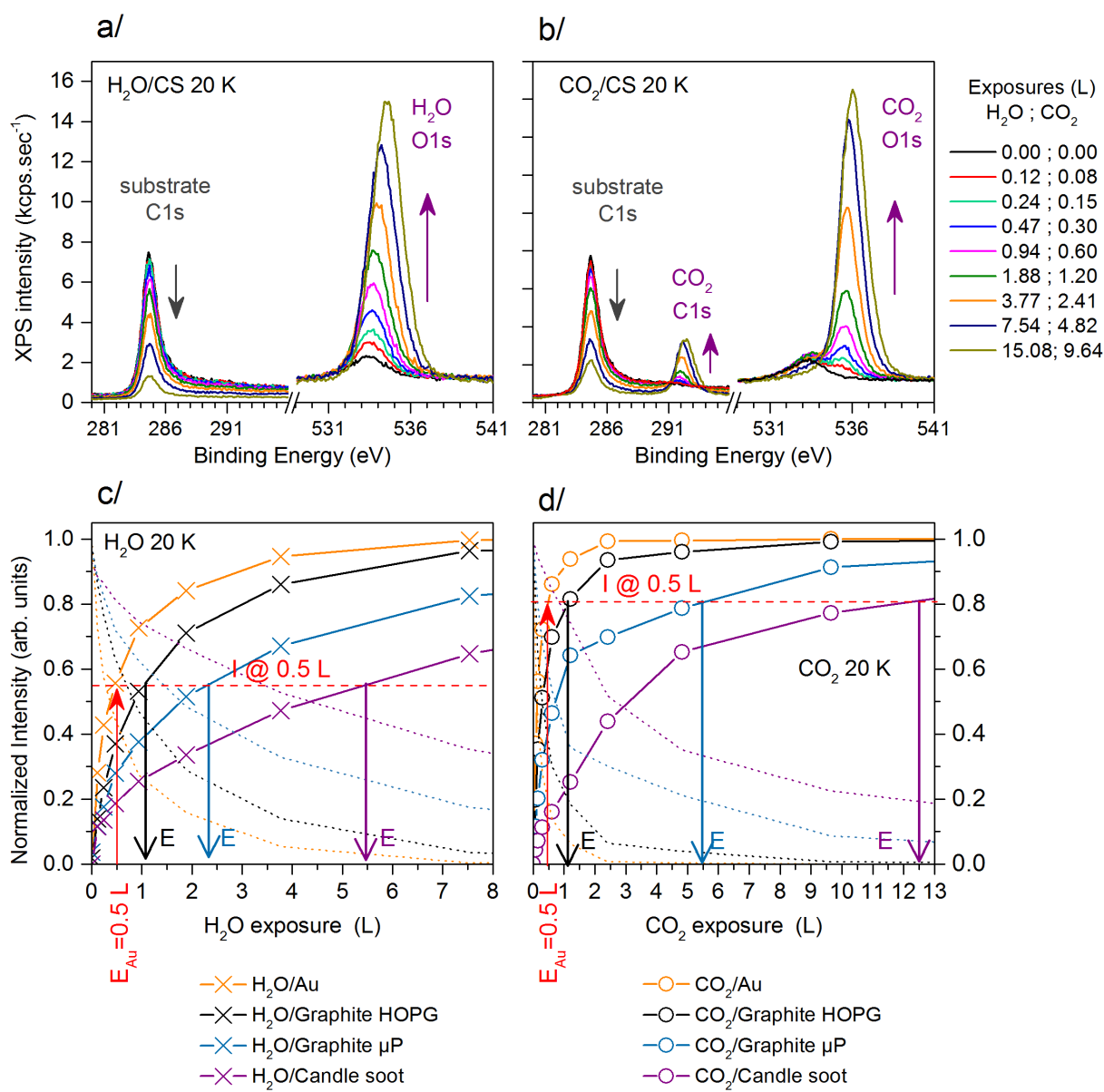

**Supplementary Figure 2** (a) evolution of the XPS signals of the substrate (C1s) and the $H_2O$ layer (O1s) as a function of the $H_2O$ and (b) $CO_2$ exposures (in L) on a candle soot (CS) film at 20 K ; (c) intensity analysis of the XPS data for $H_2O$ adsorbed at 20 K on gold, HOPG, graphite microparticles (µP) and CS; (d) same analysis for $CO_2$ at 20 K. In dotted lines are plotted the substrate intensities; in lines and symbols are plotted the adsorbate intensity. The arrows illustrate the method to extract the sticking coefficients for $H_2O$ and $CO_2$, here at 0.5 L (see text).

while the C1s signal from the substrate (~285 eV) decreases, as the adsorbed layer reduces the probability of electrons escaping from the substrate. The ∼ + 1eV shift of the O1s and C1s lines with the $H_2O$ or $CO_2$ exposures is due to the build-up of a new electronic work-function as the layers grow. A slight narrowing of the peaks (-0.2 eV) is also observed.

The fraction of the XPS intensity related to the adsorbate (or the substrate) signals in the total intensity is given by equations (S2):

$$I(\text{adsorbate}) = \frac{A(\text{adsorbate})}{[A(\text{substrate})+A(\text{adsorbate})]} \quad \text{and} \quad I(\text{substrate}) = \frac{A(\text{substrate})}{[A(\text{substrate})+A(\text{adsorbate})]} \quad \text{(S2)}$$

then

$$I(\text{adsorbate}) = 1 - I(\text{substrate}) \tag{S3}$$

Supplementary Figure 2 also presents the evolution of the XPS intensities of the adsorbate for increasing exposures of $H_2O$ (c) and $CO_2$ (d) at 20K on gold, graphite HOPG, graphite microparticles, and candle soot. In ordinates, the intensity $I$ of the adsorbate relative to the total XPS signal is calculated (in %) using equation (S4):

$$I(\text{H}_2\text{O}) = \frac{A(\text{O1s})}{[A(\text{O1s}) + A(\text{substrate})]} \text{ or } I(\text{CO}_2) = \frac{[A(\text{O1s}) + A(\text{C1s})]}{[A(\text{O1s}) + A(\text{C1s}) + A(\text{substrate})]} \tag{S4}$$

where the numerator is the peak area ($A$) of the O1s line associated to $H_2O$, or the sum of the peak areas of the O1s and C1s lines associated to $CO_2$, and the denominator is the sum of the XPS intensities of all the atomic elements of the adsorbate and those of the substrate.

These peak areas are obtained after background subtraction and peak deconvolution performed with the Casa XPS data processing software. Deconvolution also allows to distinguish the O1s and C1s adsorbates contributions from a possible O1s contribution of the substrate. This latter may be due to a substrate's faint oxygen contamination (as in the case of CS, where a small O1s peak is present prior to any $H_2O$ exposure), or because the substrate is an oxide (olivine, $TiO_2$, $Al_2O_3$), or both. For the XPS lines emitted from the bulk atoms of the substrates (e.g. C1s line of graphite HOPG, graphite µP or candle soot), the areas of the corresponding atomic emitters are corrected by their bulk relative sensitivity factor (RSF) provided by Casa XPS. For the adsorbates in the submonolayer regime, because the XPS signal is not attenuated by elastic electron scattering, the RSFs of the surface atoms differ from those of the bulk atoms, and must be corrected according to Wagner[3]. Regarding $CO_2$, we checked for each experiment that $A(\text{O1s})/A(\text{C1s})$ is equal to the expected stoichiometric ratio of 2 after these sensitivity corrections. Note that in Supplementary Figure 2 c,d the adsorbate signal saturates when the thickness of the condensed layer is significant (e.g. above 5 L of $CO_2$ on gold or graphite HOPG) and reaches a value corresponding to the maximum escape depth of the O1s electrons.

On rough or nanostructured samples, the open surface at the vacuum interface is larger than on a plane surface of the same material. Thanks to the isotropic dosing method, all surfaces of the substrate, plane or rough, have been exposed to the same molecular flux, and then the net adsorbate signal (e.g. O1s line of $H_2O$) will change according to the roughness. That is, the larger the surface area of the substrate, the larger the total amount of deposited molecules. As the normalized intensity given by equation (S4) represents the proportion of the XPS signals emitted by the adsorbate relative to those emitted by all the atoms in the analyzed area, it is independent on the surface area of the sample. This allows comparing the sticking coefficients between each sample, whatever its roughness.

When a molecule adsorbs on a part of a substrate where its electron emission is shadowed by a feature of the substrate, it does not participate to the XPS intensity. Neither did the same hidden part of the bare sample, which was equally shadowed by the same feature prior to adsorption. Equation (S4) thus only accounts of XPS signals coming from unshadowed parts. In addition, as explained below, the morphology of these unshadowed parts has no or only little effect on the determination of the amount of adsorbate, provided that the angle between the detection direction and the surface normal of the rough sample is around a "magic" angle[4]. Finally, equation (S4) also corrects from small variations in the experimental conditions, such as in the sample - X-ray source distance, or in the X-ray intensity.

The XPS peak intensities can be used to estimate the amount of material deposited on a substrate. Sophisticated methods accounting for electron inelastic (and elastic) scattering are required to simulate the whole XPS spectrum (background and peaks), or to accurately determine the thickness of the deposit on a flat or nanostructured substrate, including a nanoparticle[4,5]. However, the experimental signal can be satisfactorily approached by the conventional straight line approximation[6]. In this frame, for a flat substrate covered by an overlayer of thickness $d$, the XPS signal is attenuated with distance through the overlayer of adsorbed material according to a Beer-Lambert law. This leads to a simple exponential decay (equation (S5a)). Conversely, the signal emitted by the overlayer increases with $d$ (equation (S5b)) [7]:

$$I(substrate) \; \alpha \; e^{-d/\lambda_s cos\theta} \qquad (S5a)$$
$$I(adsorbate) \; \alpha \; (1 - e^{-d/\lambda_a cos\theta}) \qquad (S5b)$$

where $\lambda_s$ is the inelastic mean free path (IMFP) of electrons emitted by the substrate and moving in the adsorbate, $\lambda_a$ is the IMFP of electrons emitted by the adsorbate and moving in the adsorbate layer, and $\theta$ the angle between the macroscopic surface normal of the sample and the XPS detector axis ("take-off" angle, here $\theta = 55°$).

The take-off angle $\theta$ is a key parameter to relate the XPS intensities to the thickness. When the morphology of the sample is not flat, the Beer-Lambert law (Eqs. S5) is still valid, but $\theta$ is no more univocally defined. Unless all take-off angles are integrated experimentally, for instance using an immersion lens[8], it becomes impossible to derive the thickness of the overlayer on a rough surface from equations (S5). This issue has been addressed extensively in the literature on XPS and Auger spectroscopies. It has been shown theoretically and experimentally that around a certain take-off angle, called the "magic angle", the influence of roughness is almost null, and the thickness of the overlayer can be determined accurately with an error less than 10 %[8–12]. Yet, there is no universal magic angle valid for any type of roughness[7]. For roughness that can be modelled by spherical segments, as for the film of nanoparticles studied here, the magic angle lies around 55°[12]. We thus used this value. We assume that the largest error due to roughness will occur for the roughest samples, which is likely the p-ASW ice deposited on a carbon nanoparticles film. This system combines two roughnesses, that of the nanoparticles and that of the porous ice layer.

One can experimentally validate the assumed value the magic angle $\theta_m$. On a rough surface observed at this angle, if the influence of roughness is null, the effective distance through which the electrons pass – hence the corresponding XPS intensity - is equal to that of a flat overlayer of thickness $d$, which is $d/cos\theta_m$. Then, the magic angle can be checked as follows[12]. In Supplementary Figure 3 is

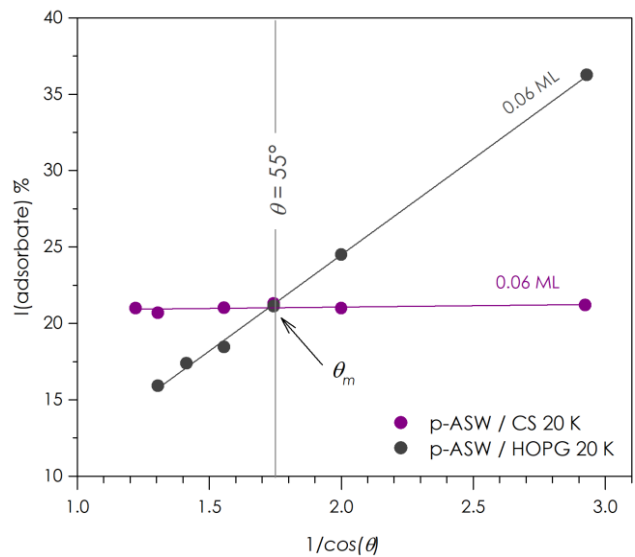

**Supplementary Figure 3** angular evolution (plotted as a function of $1/cos\theta$) of the intensity of XPS signal recorded on a submonolayer of p-ASW ice adsorbed at 20 K on a candle soot film and on HOPG at a coverage of 0.06 ML. The vertical lines indicate the take-off angle used in the experiment (55°). The intensities emitted by the ice layer deposited on HOPG and on the soot film are identical at 55°, which defines the "magic angle" $\theta_m$ where the influence of roughness on the signal is null.

plotted the angular dependence of the XPS intensity of the adsorbate as function of $1/cos\theta$ for a submonolayer (0.06 ML) p-ASW ice film deposited at 20 K on a candle soot film, compared to the angular dependence for a same film deposited on a flat HOPG substrate at the same coverage. On HOPG, the adsorbate signal varies as expected, *i.e.* linearly with the take-off angle, while it remains constant on the CS substrate, as also expected for spherical nanoparticles[12]. The two signals intercept at the 1/cosθ value corresponding well to the assumed magic angle. We did not checked the magic angle for every substrates, but it might be valid for a substrate of hemispherical particles having a wide range of diameters (60-490 nm), and can apply in practice to more complex surface morphologies[12]. We assume then for all the particles studied here, the error due to roughness remains within the admitted error of 10%.

Let us now explain how are extracted graphically the sticking coefficients. In equations (S5), d is the amount of deposited material per unit area; it is equal to the molecular dimension $l$ (i.e. the thickness of an ice monolayer) times the coverage, which is the sticking coefficient S times the exposure E:

$$d = l.S.E \qquad (S6)$$

It can be seen in Supplementary Figure 2 c,d that when dosing $H_2O$ or $CO_2$ on graphite HOPG, graphite µP and on candle soot, larger exposures $E$ are required to achieve XPS intensities equal to that obtained when dosing on gold. For a given substrate, achieving the same XPS intensity of the adsorbate layer than on gold (*i.e.* the same amount of adsorbate) satisfies the following equality (using equations (S5b) and (S6)):

$$1 - e^{-l.S_{Au}.E_{Au}/\lambda_a} = 1 - e^{-l.S.E/\lambda_a} \qquad (S7)$$

Leading to:

$$S_{Au}.E_{Au} = S.E \qquad (S8)$$

As $S_{Au} = 1$ by calibration, the sticking coefficient of a given substrate is simply given by the ratio of the exposures:

$$S = E_{Au}/E \qquad (S9)$$

For instance on the adsorbate intensity, at a given exposure of 0.5 L of $H_2O$ or $CO_2$ on gold ($E_{Au} = 0.5$ L) corresponds a specific intensity, indicated by a horizontal dashed line in Supplementary Figure 2 c,d (55 % for $H_2O$, 70 % for $CO_2$). The intercepts of this line with the experimental curves obtained on the different substrates provide the corresponding exposures $E$ (given by abscissae indicated by the vertical arrows) required to achieve the XPS intensity obtained on gold, which provides - at this exposure- the sticking coefficients using equation (S9).

Since the goal of this work is the study of the sticking coefficient on the bare surfaces, we have made this graphical analysis for submonolayer exposures of 0.1, 0.25, 0.5, 0.75, and 1.0 L. Table 1 of the main text presents the sticking coefficients of $H_2O$ and $CO_2$ determined for these exposures on the bare substrates at 20 K, and those of $CO_2$ at 20 K on substrates pre-covered with porous (p-ASW) and non-porous (ASW) amorphous ice. The values of S discussed and presented in the Fig. 2 and 3 of main text correspond to their average (right column in Table 1). In some cases, S values change with exposure, especially on flat substrates. This results from the evolution of the growth mode with coverage, from isolated adsorption at low exposures, which is expected to dominate because the probability of two adsorption events at the same site is low, to the formation of 2D islands or 3D multilayers patches at higher exposures. However, on nanoparticles the sticking coefficients remain low and fairly constant, indicating that, contrary to flat surfaces, the growth mode does not change significantly with exposure. Because of the low sticking probabilities on these particles, the coverages remain small below 1 L of exposure and, due to the lack of material, adsorption is more likely to occur in 1D isolated mode and in 2D island formation.

### Supplementary Information 3. Error estimation on S

1-Due to different slopes of the curves given Supplementary Figure 2 c-d, an error on exposure will lead to an error depending on the substrate. The errors on the XPS intensity have been determined by the spread resulting from an error of 10% on the exposure (gauge accuracy, rise and fall fronts during dosing). For this, we chose the exposure of 0.5 L, in between 0 and 1 L, the useful range for S determination. It leads to errors of:

±2.5-3.5% on nanoparticles
±1.5-2.5 % on graphite microparticles
±1-1.5 % on flat samples

The lower range corresponds to errors for $H_2O$, the upper range corresponds to error for $CO_2$, since the XPS intensities are smaller due to smaller S. There is not significant difference when these surfaces are covered with ice, so these errors are the same for the sticking of $CO_2$ on the corresponding bare surfaces.

2-The detection angle of 55°, while it cancels roughness effects for soot (for which error due to roughness is thus 0%), might be not perfectly suited for other samples, which might add an error of around ±5% [10,12] :

0% for soot
±5% for $Al_2O_3$ microparticles, $TiO_2$ microparticles and Graphite microparticles samples
0% for the flat surfaces for which roughness is considered negligible.

3-We use a commonly accepted value of ±5% for the quantitative accuracy of the XPS analysis (background removing, fitting procedure, accuracy of the RSFs, etc.).

Errors 1, 2 and 3 being uncorrelated, they add in quadrature, and are the standard deviation (s.d.) of the measurements.

- Finally, since the graphical determination of S is the ratio of the XPS intensities measured on the gold reference with that of each substrate, errors on S are the quadratic additions of the total error on the gold reference (±5%) with that of the considered substrate and molecule.

| Substrate | Error on S($H_2O$) | Error on S($CO_2$) |
|---|---|---|
| Flat surfaces | ±7% | ±8% |
| Graphite microparticles | ±9% | ±9% |
| $Al_2O_3$ , $TiO_2$ and olivine microparticles | ±9% | ±10% |
| Soot | ±8% | ±8% |

### Supplementary Information 4. XPS thermal desorption experiments

Any temperature gradient between the back of the sample and the exposed surface must be avoided. Otherwise the sticking coefficient will be measured in a temperature range higher than that set, and could then vary greatly[15]. We therefore systematically verified that $H_2O$ desorption occurred in the expected temperature range, around 140 K. On all substrates, we measured the desorption temperature $T_d$ of a $H_2O$ ice submonolayer coverage by monitoring the O1s and substrate's XPS lines while annealing from 20 K to 200 K (heating rate of $1 \text{ K} \cdot \text{min}^{-1}$). Examples of experimental curves are presented in Supplementary Figure 4, zoomed in on the temperature range in which desorption occurs. These curves at normalized to 1 to correct for intensity variations resulting of the different exposures from a substrate to another. $T_d$ was systematically found in the expected temperature range, although it might slightly vary from a substrate to another due to differences in interaction energies with the substrates, and within molecules in the condensed layer[16]. Also note that no change in sample temperature was detected during the exposures.

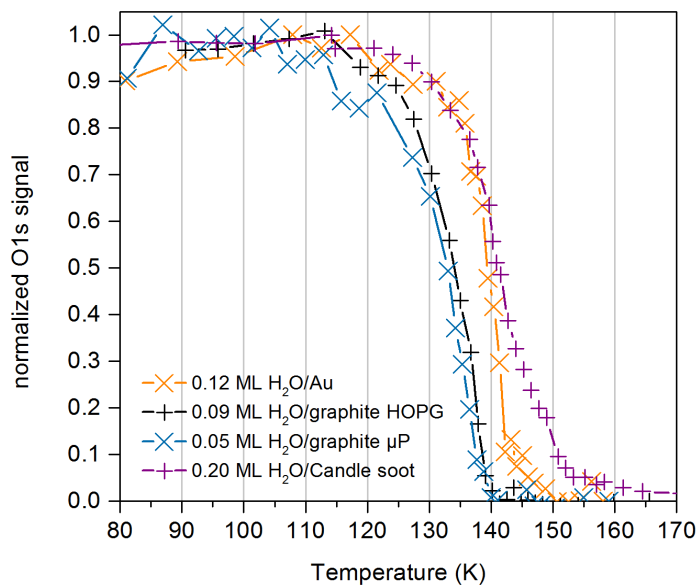

**Supplementary Figure 4** Evolution of the O1s signal of $H_2O$ ice deposited at submonolayer coverages (indicated in the legend) on a selection of substrates (Au, HOPG, graphite μP and candle soot), during thermal desorption at a heating rate of $1 \, K \cdot min^{-1}$.

These results also show that the desorption temperature of $H_2O$ - and therefore its adsorption energy - is almost identical whatever the surface considered, and whether the probability of sticking is high (Au) or very low (CS).

### Supplementary Information 5. Dependence of S on the temperature

When adsorption is carried out at a substrate temperature Ts in the vicinity of the desorption temperature Td, *e.g.* 140 K for $H_2O$, the probability for a molecule to desorb back to the vacuum increases as Ts approaches Td[14]. However, when adsorption is carried out at Ts <<Td, as in this study, we expect no or only slight temperature dependence of S. As shown Supplementary Figure 5, no significant difference in the sticking coefficient of $H_2O$ is observed between 20 K (blue) and 80 K (red).

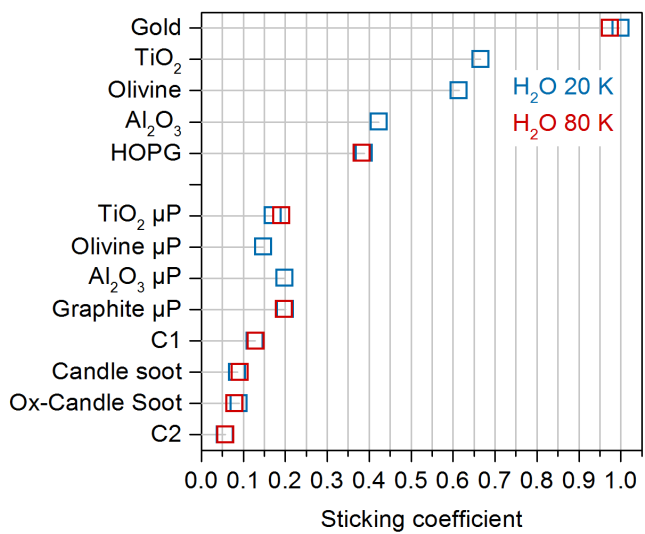

**Supplementary Figure 5** Sticking coefficients of $H_2O$ on various substrates at 20K (blue) and 80 K (red).